\documentstyle[prl,aps,epsf]{revtex}

\begin{document}
\begin{twocolumn}
\draft
\tolerance 50000

\title{Less singular quasicrystals: life in low codimensions }
\author{A. Jagannathan}
\address{ Laboratoire de Physique des Solides,
Universit\'{e} Paris-Sud,
91405 Orsay, France \\
Department of Physics and Astronomy, University of Southern California, Los Angeles, USA}
\maketitle
\vskip 2cm
\begin{abstract}
We consider a set of tilings proposed recently as d-dimensional 
generalisations of the Fibonacci chain,
 by Vidal and Mosseri. These tilings have 
a particularly simple theoretical description, making them 
appealing candidates for analytical solutions for electronic
properties. Given their self-similar geometry, one could expect that
the tightbinding spectra of these tilings might possess the 
characteristically singular
 features of well known quasiperiodic systems such as the 
Penrose or the octagonal tilings. We show here, by a numerical study
of statistical properties of the tight-binding spectra 
that these tilings fall rather in an 
intermediate category between the crystal and the quasicrystal, $i.e.$ in a
class of almost integrable models. This is certainly a consequence of
 the low codimension
of the new tilings.
\end{abstract}

\section*{Introduction}
We report the results of a numerical study of the two-dimensional (2D)
members of a new family of
quasiperiodic tilings proposed recently by Vidal and Mosseri 
\cite{vid}. 
These new so-called generalized
 Rauzy tilings (GRT) (after the work of \cite{rau}), 
are generalizations of the
Fibonacci chain and are simpler in structure than hitherto studied
tilings in two and three dimensions. Tight-binding models on these systems
may be amenable  to analytical techniques
whereas other 2D tilings thus far studied have been refractory to
approaches including renormalization group, or recursive methods
\cite{moss,aj2}. This property makes the new tilings attractive candidates to
study in the world of quasiperiodic Hamiltonians.
It is thus interesting to examine whether the tight-binding spectrum
of these new tilings have similarly complex structures as for the 
``canonical" cases studied earlier, namely the Penrose or the Ammann-Beenkker
 (octagonal) tilings. This turns out not to be the case: 
 these tilings
fall in a class intermediate
between the periodic lattice, an integrable problem, and the
octagonal tiling, which belongs in the non-integrable class.

This conclusion is based on 
calculations of statistical properties of the energy
levels of periodic approximants of these new systems.
Such analyses have been carried out
for complex systems, including disordered media,
systems with interactions 
or quantum billiards \cite{and}. 
 A relation is found to exist between the quantum statistics and the
classical dynamics : depending on whether the dynamics is
  classically ergodic, or classically
chaotic, one finds their quantum properties
to be described by Poisson statistics or by 
Random Matrix Theory (RMT) \cite{rmt}.
 Such  
statistical analyses are useful for gaining insight into
 complex systems for which exact, 
or even approximate theoretical results are not available. This is the case
 for the octagonal tiling, a quasiperiodic tiling in two dimensions which
has been tackled via numerical study of tight-binding models 
\cite{bs,pj,zh}.
We show that the GRT in two dimensions has spectral
properties that differ $essentially$ from those of the octagonal
tiling. The GRT appears analogous to the 
pseudointegrable billiards proposed in \cite{bog} which are neither 
completely integrable nor completely chaotic.


\section*{Description of the new tilings}
The cut-and-project method provides a simple and easily implemented 
way to obtain pieces of quasiperiodic tilings. One can thus obtain
arbitrarily large pieces of an infinite quasicrystal, or one can construct
 finite pieces that can be periodically continued -- these  are the
periodic approximants of the quasicrystal.  
As described in \cite{dun}, one 
first defines the real-space directions, which constitute a d-dimensional
subspace of a larger D-dimensional hyperspace. The remaining D-d directions
constitute the `perpendicular" space. The quasicrystal is obtained by 
projecting all the points contained within a certain infinite strip
in the hyperspace \cite{strip}
onto the real space. 

When  
the orientation of the real space is irrational, 
the tiling is aperiodic, while if rational orientations are chosen,
the tiling repeats periodically in
real space. This 
method has been used to obtain the Fibonacci chain (D=2, d=1), the
octagonal tiling (D=4, d=2), 
the three-dimensional
 Penrose tiling (D=6, d=3) etc. The codimension, D-d, is in fact
equal to d in the three cited cases. 

The simplifying feature of the GRT 
\cite{vid} is that their codimension is equal to 1 for all
real space dimension d. For a finite tiling, this allows one to
 index the points  according to their perpendicular space
coordinate.
Furthermore, when the points of an approximant of these tilings -- 
illustrated
for two dimensions in Fig.1 --
 are so
indexed, the 
connectivity matrix of the approximant (assuming 
toroidal boundary conditions) takes on a very simple 
band-diagonal (Toeplitz) structure \cite{vid}.
 In the two dimensional case that we consider here, the nonzero matrix elements
are situated at distances of $F_{n-3}, F_{n-2}$ and
$F_{n-1}$ from the diagonal. The 
generalized Fibonacci numbers $\{ F_{n}\}$ are 
obtained from a three term recursion relation $F_{n} = F_{n-1}
+ F_{n-2} + F_{n-3}$, with the 
initial conditions $F_{-1} = 0; F_{0} = F_{1} =1$. 
The ratio $F_{n}/F_{n-1}$ tends to the value
 $\alpha \approx 1.839...$, solution of the equation
$x^3 = x^2 + x + 1$. 
These matrices correspond to approximants of increasing 
size as $n$ is increased. 

The model we consider is a pure hopping tight-binding 
Hamiltonian, 
\begin{equation}  
 H = \Sigma_{\langle i,j \rangle} t c^{+}_{i} c_{j} + h.c.
\end{equation} 

 where $i,j$ in the sum correspond to pairs of sites linked by a bond in
Fig.1. The hopping amplitude has been
 assumed to be independent of the linked sites
 and is set to unity.
  We consider the case of periodic boundary conditions,
and take $k=0$ where $k$ is
the Bloch vector.
The resulting eigenvalue problem reduces to that of
  diagonalizing the connectivity matrix 
 defined above. This was done using a Lanczos routine. The spectrum is 
 expected to be symmetric for the infinite tiling, since it is 
 bipartite. In addition, there is a 
 discrete symmetry under an inversion in the perpendicular subspace \cite{inv}.
 This results in two subspaces of positive and negative parity 
 respectively, and our analyses are carried out for each sector separately.

\section*{Results}
 The spectral density $\rho(E)$ is shown in Fig.2a, with that of the octagonal 
 tiling shown in 2b for comparison (the localized states of the latter at $E=0$ have not been included in the figure). 
The density of states curves shown were obtained for  given sample sizes
assuming periodic boundary conditions, and illustrate the characteristic
rapidly fluctuating behavior typical of these aperiodic tilings (similar 
curves were found in \cite{vid} and \cite{bs}). It is evident that 
while the GRT 
 has an underlying smoothly varying component of $\rho(E)$ which is
everywhere non-zero (notice the y-axis offset) whereas
 the octagonal tiling does not. Calculations on approximants of the
 2D Penrose tiling show strong fluctuations as well with a
multitude of gaps and pseudogaps \cite{janss}.

  The envelope of $\rho(E)$ of the GRT is in fact reminiscent of 
 the density of states of the square lattice. One notes in Fig.1 that there are
many regions in which sites of coordination number 4 are grouped together in
this tiling. In contrast to the octagonal tiling, which also has a mean coordination
number $\overline{z} =4$, the fluctuations of geometry are less strong in
the GRT. This is a result of the reduced codimension, and we see
that in consequence the density of states has a much less singular structure
as compared to the octagonal case.

Turning to the statistics of nearest neighbor level spacings, we investigate
the distribution of $E_{i+1} -E_i$ where the $E_i$ are the ordered set of 
energy levels of the Hamiltonian (1). In order to eliminate the trivial
dependence of this quantity on variations of $\rho(E)$, one first carries out
an ``unfolding" of the spectrum \cite{boh} to get the corrected level spacings
\cite{unfol}. This yields the unfolded spacings $s_i$ whose mean value is
equal to 1.
The resulting distribution of spacings $P(s)$
 is plotted in Fig.3a for the three largest systems studied, which contain
 35738(L), 19513(M) and
10609(S) sites.
The dashed line represents the Poisson $\rm{e}^{-s}$ decay, while
 the continuous line is the semi-Poisson form $P_{SP}(s) =4s \exp (-2s)$. The latter
has been proposed for pseudointegrable billiards
 and can be shown to correspond to a certain short range plasma
model (SRPM) by Bogomolny et al \cite{bog}. 
 These authors speculate that this
type of statistics is a consequence of fractal structure of the wave functions
in those systems. In the Anderson model, the
critical level statistics were first shown to have the semi-Poisson distribution by
Braun et al \cite{gill}. In another example of this universality class 
 Evangelou and Pichard had considered the distribution of bandwidths and
obtained the semi-Poisson form in the critical Harper model \cite{eva}.

In Fig.3a one sees that there is a size-dependence in the fall-off
 in $P(s)$ at small $s$.
 For the largest size, the points appear to 
deviate substantially from the semi-Poisson curve. It will be interesting to 
calculate the next system size to see if the deviations are
systematically upwards, i.e. towards the Poisson limit. The $P(s)$
resembles that obtained in \cite{gill} for periodic boundary conditions at the
critical point of the Anderson model. 
 In contrast to the random case however,
 in our case there is a 
size-dependence. It is important to note that
the $degree$ of unfolding matters here:
 nonunfolded spacings follow a simple Poisson law while 
progressively stronger unfolding causes the distribution to go over to the
semi-Poisson form. In Fig.3b we present a semilog plot of the
statistics of the maximally unfolded level spacings ( where we
renormalized each spacing by the locally averaged value over the nearest
neighbor energy levels). The agreement with the semi-Poisson form is seen to 
be good over a wide range of values. At large $s$ $P(s)$ falls off faster
than exponentially in $s$, while at small $s$, there is a size dependent increase
of $P(s)$ above the semi-Poisson value.
We believe that these effects may be explained by the unfolding
procedure, although boundary condition effects cannot be ruled out and further
studies are needed to settle this issue.

This distribution in Fig.3a is very different from the one obtained for
  the octagonal tiling. For that tiling, unfolded spacings 
 follow the Wigner-Dyson distribution for 
 random matrices of the gaussian orthogonal ensemble (GOE),
$P_{GOE}(s) = \frac{\pi}{2} \exp(-\pi s^2/4)$. 
If one considers the distribution of the ``bare" spacings (i.e. without
correcting for the variation in density of states), they follow
a broad (log-normal) distribution law :
 this can be shown via a recursive
relation based on the inflation symmetry of the tilings \cite{aj}. A less
singular case is that of a
randomized version of the octagonal
tiling \cite{pj}. 
The randomized tiling has a less fluctuating density of states
and the spacings distribution is simply GOE (unfolding does not produce
any new effects).
From  these $P(s)$, as
 well as from the spectral rigidity function (to be discussed 
 below) we conclude that the octagonal tilings
  belong in the class of non-integrable 
 models exemplified by disordered metals. One has an unbounded 
 diffusion of wave packets resulting in level repulsion at small $s$. 
 The GRT is clearly not in the GOE regime for any scale 
 of energies, rather it
 should be considered a nearly integrable system, as the shape of 
  $P(s)$ indicates.

 This becomes more evident upon looking at the spectral rigidity function 
 $\Sigma^{(2)}(L) = \langle (N(L) - \overline{N})^{2}\rangle$, where
$N(L)$ is the number of levels lying within an energy interval of width $L$.
The angular brackets mean averages over different starting positions of the 
interval within the spectrum. 
Fig.4a shows the plot of $\Sigma^{(2)}$ for the largest tiling size, for
three different choices of unfolding \cite{unfol}
 The continuous line in Fig.4a indicates
the SRPM result $\Sigma^{(2)}(L) = \frac{1}{2}(
L + (1 - {\rm e}^{-4L})/4)$. For very small $L$, the dependence
appears well described by the semi-Poisson law.
The three curves have the same low
energy behavior, but the ``most unfolded" case has its fluctuations
damped out because of the unfolding, while the other cases show the
transition from approximately linear to approximately quadratic
dependence on $L$.

The SRPM behavior at small $L$ may be related to a long time dynamics
in this tiling which is intermediate between chaotic and integrable. This
would correspond to the semi-Poisson form of $P(s)$ 
It is interesting to consider the rigidity
evaluated for the energy levels without unfolding.
In the case of random
media, where diagrammatic perturbation theory can be exploited, and
checked by more semiclassical methods \cite{alt}, it was shown that in the 
  diffusive regime, as a function of energy $E$,
 $\Sigma^{(2)}(E)$ has two different sorts of 
 behavior: firstly, at very low energies, it has the logarithmic 
 dependence expected for GOE matrices. This holds
 upto the scale of the Thouless energy
 $E_{c}$ \cite{thou}
which is size dependent, corresponding to
the time taken to diffuse to the boundaries. For larger energies
  $\Sigma^{(2)} \sim E^{d\nu}$ 
 where the exponent $\nu$ describes the spreading  over time
 of an initially localized
 wave packet  $\langle r^2\rangle
 \sim t^{2\nu}$. This theoretical prediction was
verified numerically in the Anderson model \cite{dan}, where the exponent
is one for ordinary diffusion, $\nu = 0.5$.
For the randomized 
 octagonal tiling, 
  $\nu\approx 0.85$ \cite{pj} as deduced from $\Sigma^{(2)}$,
 in  agreement
 with the results of a direct calculation of wavepacket
diffusion on this tiling \cite{bps2}. 
 Fig.4b shows the log-log plot of $\Sigma^{(2)}$ as a function of energy 
 for three sizes. There is a crossover from linear dependence
to an almost quadratic
dependence, $\Sigma^{(2)}(E) \propto E^{2\nu}$ ($\nu \approx 0.95$). 
The sub-quadratic regime
 very likely corresponds to the sub-ballistic regime of dynamics
that was observed for wavepackets diffusing on
a 2D GRT by Vidal and Mosseri \cite{vid2}.

\section*{Discussion and Conclusions}
The two-dimensional GRT tiling, despite its quasiperiodic structure, 
is an almost integrable
system, with a semi-Poisson form for the level spacings distribution and
a corresponding small-energy spectral rigidity function. 
 The $P(s)$ distribution is probably linked to the existence of
fractal wavefunctions, as has already been observed in critical
quantum systems.
 In the GRT studied here, it is noteworthy that although 
 fluctuations of the density of states are present, they are strongly
reduced with respect to those in ``canonical" quasiperiodic tilings.
 This is a consequence of the reduced codimension of the
tilings, which leads to its having weaker quasiperiodic modulations of geometry
than the previously studied octagonal and Penrose tilings.
Further study is needed to elucidate the nature of states 
and of the relation between dynamical evolution of wavepackets and the
calculated statistical properties of the energy levels.

\section*{Acknowledgments}
I would like to acknowledge useful discussions with Frederic Piechon and
Michel Duneau. I thank the computing center at IDRIS, Orsay for the
computing facilities and support needed for this study.
 

\begin{thebibliography}{15}
\bibitem{strip} The strip is defined by the set of points 
whose perpendicular space coordinates lie within a finite 
``acceptance domain".
\bibitem{inv} The Hamiltonian is invariant under a reindexation of points
given by inversion of the perpendicular space coordinate with
respect to the midpoint of the strip. This symmetry
is not exactly obeyed in the approximants, however the symmetry breaking
gets weaker as system size is increased.
\bibitem{unfol}
The smoothed integrated density of states is calculated using
 a spline fit on a subset of the data points. The more points used 
in the smoothing, the greater the ``unfolding".
\end{thebibliography}

\end{twocolumn}

\newpage
\begin{figure}[h]
\centerline{\epsfxsize=90mm
\epsffile{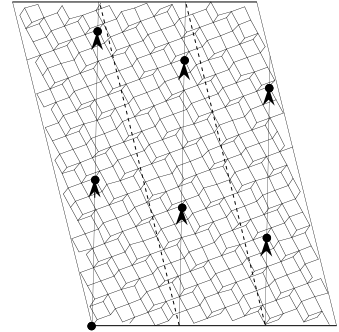}}
\vspace{-60mm}
\caption{An approximant of the 2D rauzy tiling showing the numbering
scheme: the arrows show the diplacement vector from site 0(
origin) to site 1, etc. Note the periodic boundary conditions.}
\label{approx}
\end{figure}

\vskip3cm
\begin{figure}[h]
\centerline{\epsfxsize=100mm
\epsffile{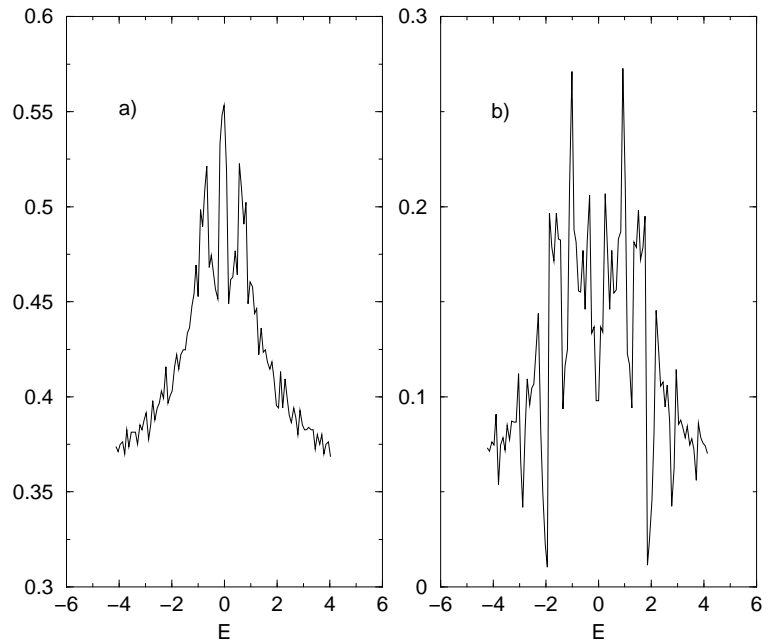}}
\caption{(a) Density of states of the 2D Rauzy tiling and (b) of the
octagonal tiling }
\label{density}
\end{figure}

\begin{figure}[h]
\centerline{\epsfxsize=85mm
\epsffile{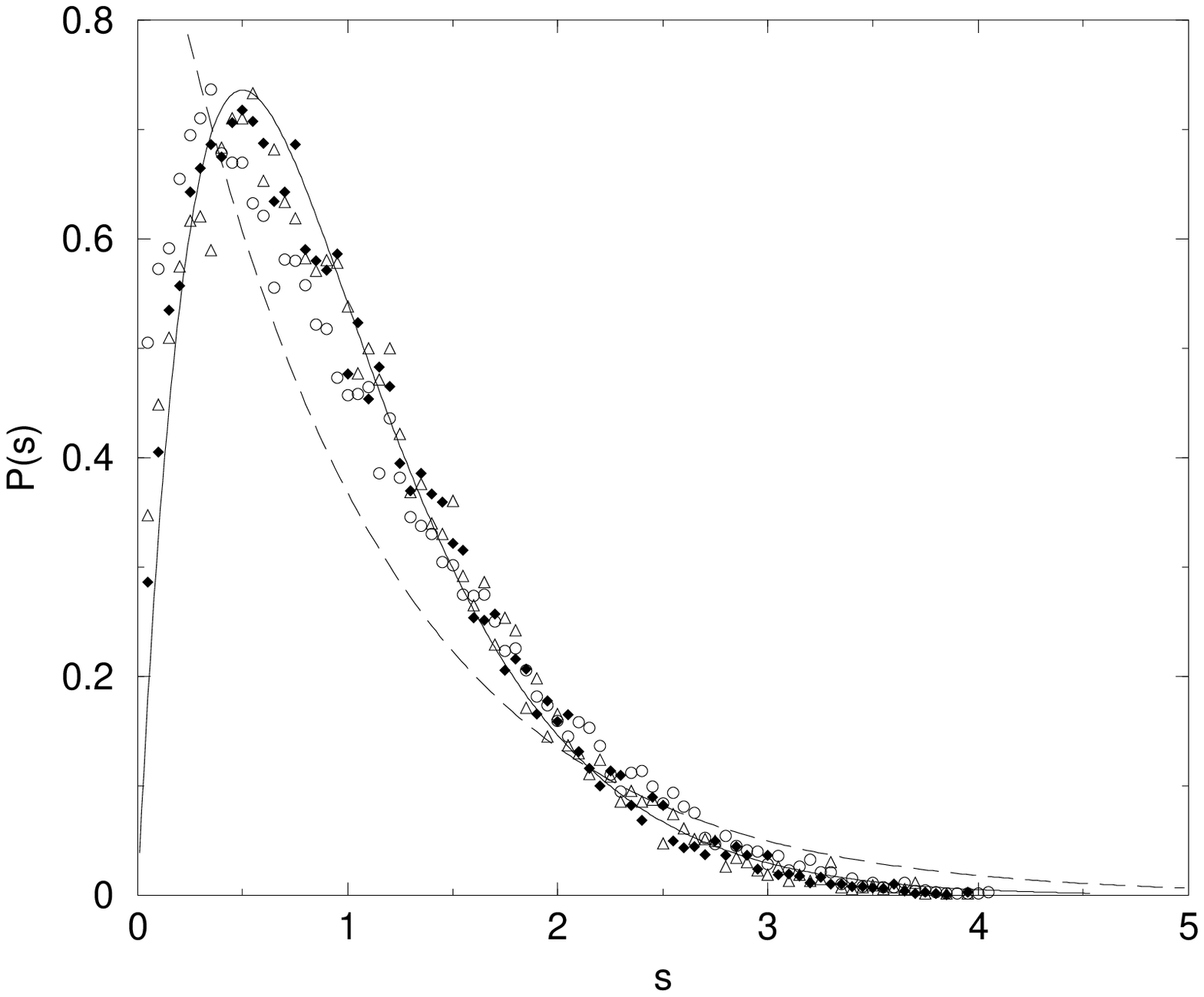}\epsfxsize=85mm\epsffile{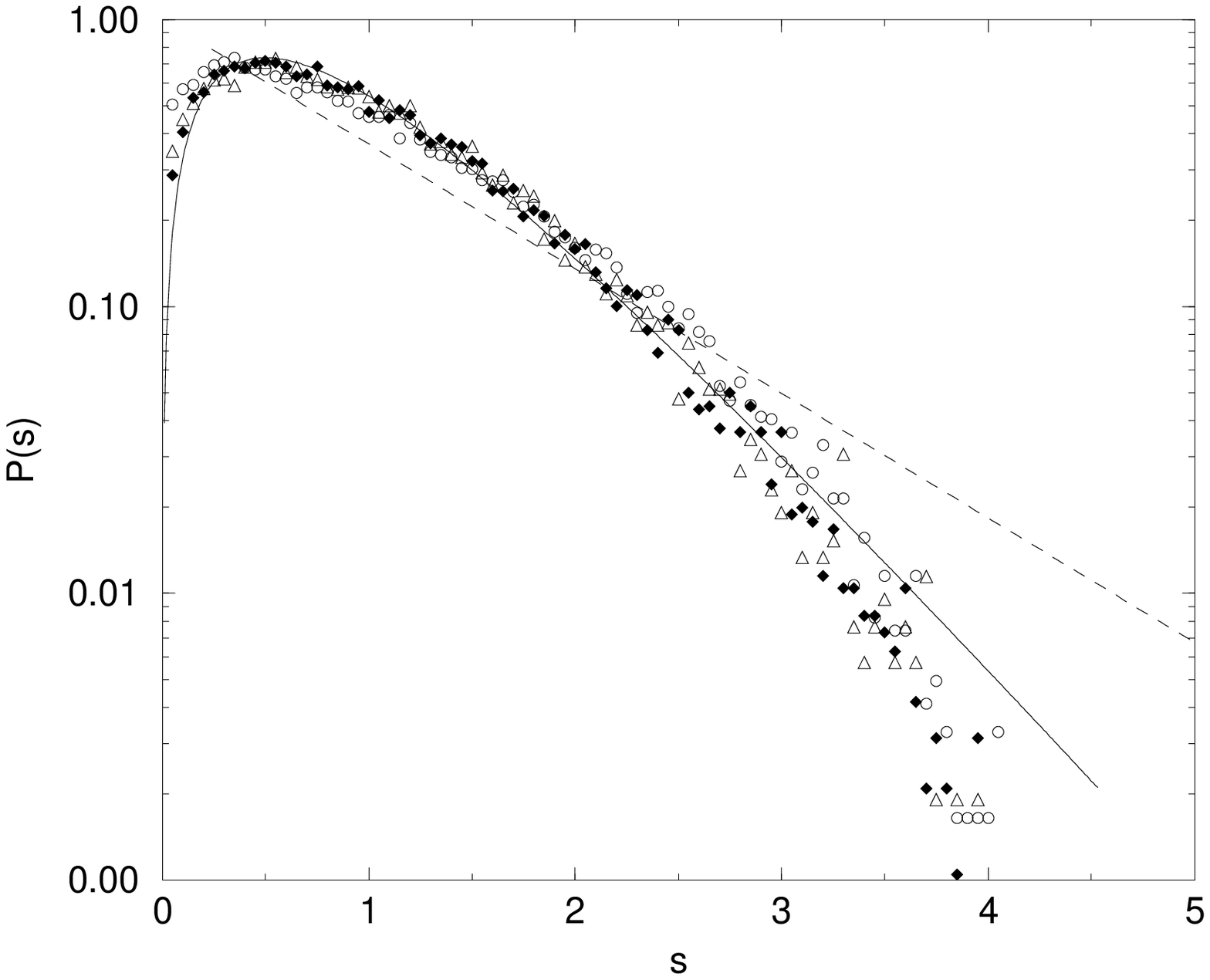}}
\caption{(a) P(s) of three approximants: L(circles),
M(diamonds) and S(triangles) along with the semi-Poisson 
(continuous line) and the Poisson (dashed line) laws.
 (b) Semilog plots of P(S) with the Poisson and the semi-Poisson
laws (dashed and continuous lines respectively).}
\label{Pofs}
\end{figure}

\vskip3cm

\begin{figure}[h]
\centerline{\epsfxsize=85mm
\epsffile{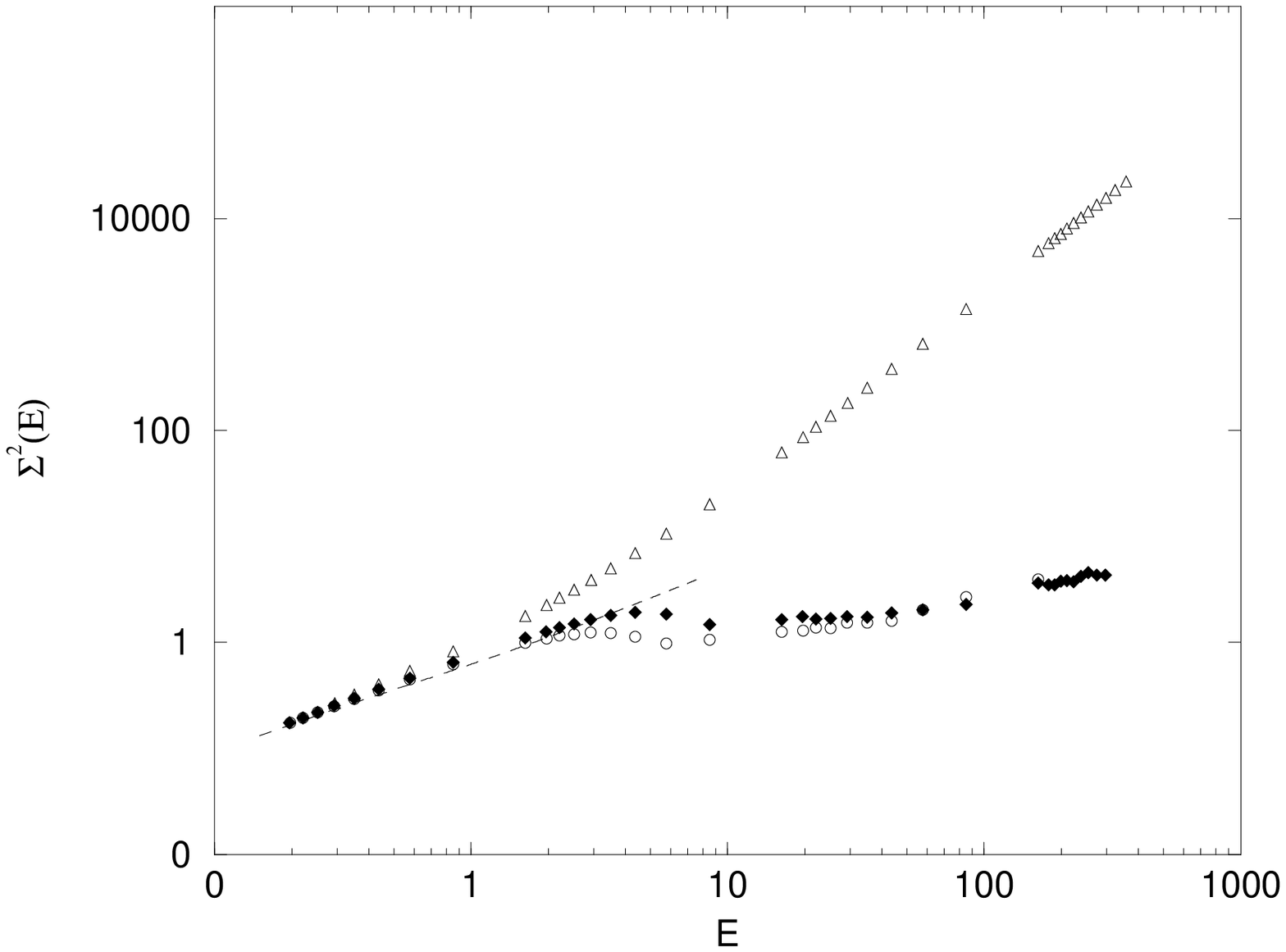}\epsfxsize=85mm\epsffile{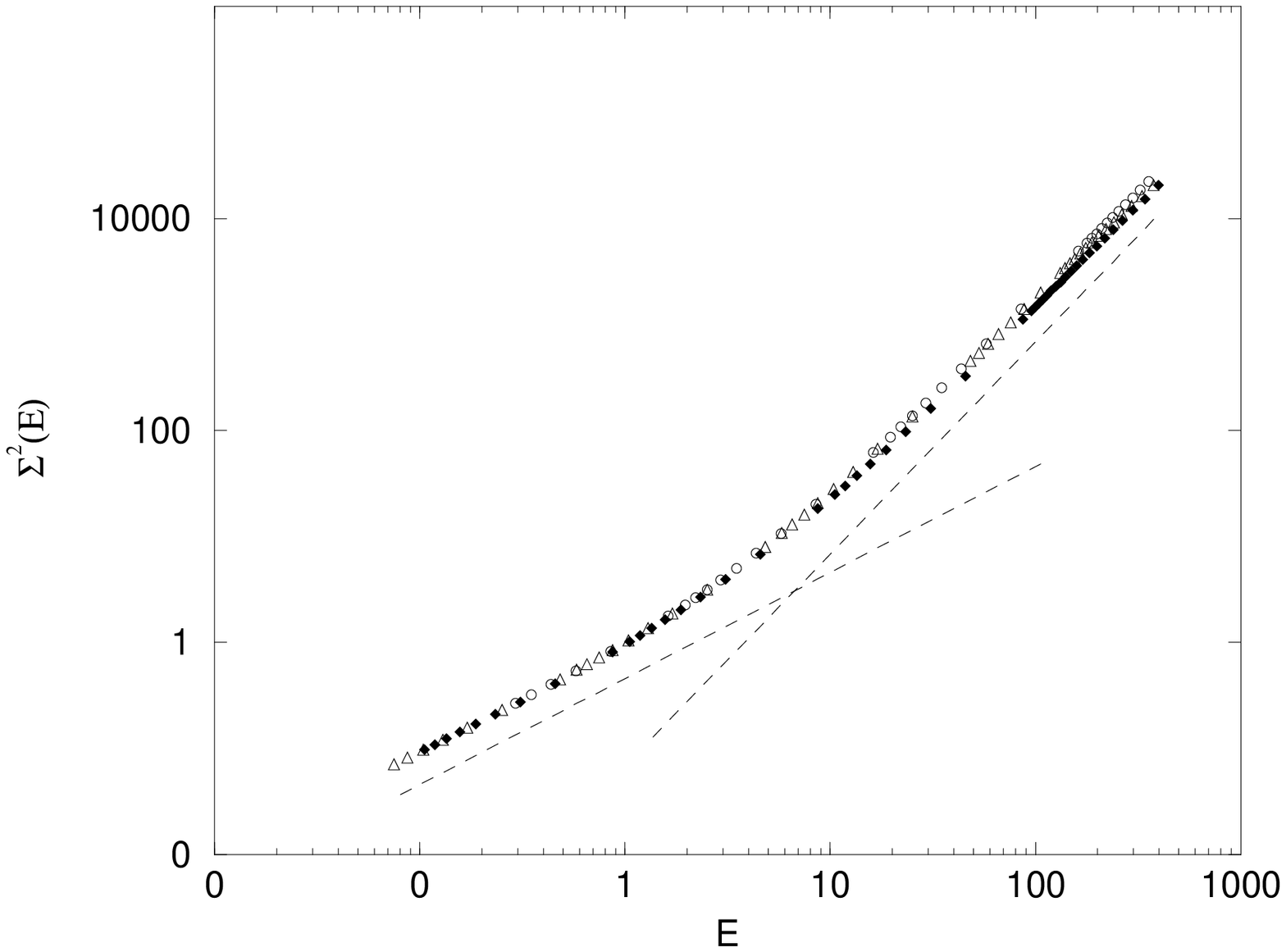}}
\caption{(a) The rigidity $\Sigma^2(E)$ for different levels of unfolding,
(see text) from most unfolded (circles) to not unfolded (triangles).
(b) $\Sigma^2(E)$ without unfolding for three sizes L,M and S. The dashed
lines show $E$ and $E^2$ power laws.}
\label{Ps}
\end{figure}
\end{document}